\documentclass[sigconf]{acmart}

\def\BibTeX{{\rm B\kern-.05em{\sc i\kern-.025em b}\kern-.08emT\kern-.1667em\lower.7ex\hbox{E}\kern-.125emX}}

\usepackage{booktabs} 
\usepackage{multirow}
\usepackage{siunitx}
\usepackage[most]{tcolorbox}
\usepackage[export]{adjustbox}
\usepackage{hyperref}
\usepackage{subfig}
\usepackage{algorithm, algpseudocode}
\usepackage{hyperref}
\usepackage{placeins}
\usepackage{amsmath,amsfonts}
\usepackage{graphicx}
\newcommand{\R}{\mathbb{R}} 

\acmConference[ESEC/FSE 2022]{The 30th ACM Joint European Software Engineering Conference and Symposium on the Foundations of Software Engineering}{14 - 18 November, 2022}{Singapore}
\acmYear{2022}
\copyrightyear{2022}

\acmPrice{xx.00}

\begin{document}

\title{Testing of Machine Learning Models with Limited Samples: An Industrial Vacuum Pumping Application}

\author{Ayan Chatterjee}
\orcid{0000-0002-8016-0965}
\affiliation{%
  \institution{Department of Mathematics and Computer Science, Karlstad University, Universitetsgatan 2, Karlstad, 651 88, Sweden}
}
\email{ayan.chatterjee@kau.se}

\author{Bestoun S. Ahmed}
\orcid{1234-5678-9012}
\affiliation{%
  \institution{Department of Mathematics and Computer Science, Karlstad University, Universitetsgatan 2, Karlstad, 651 88, Sweden}
}
\email{bestoun@kau.se}

\author{Erik Hallin}
\affiliation{%
  \institution{Uddeholms AB, Uvedsv{\"a}gen, Hagfors, 683 33, V{\"a}rmlands l{\"a}n, Sweden}
}
\email{erik.hallin@uddeholm.com}

\author{Anton Engman}
\affiliation{%
  \institution{Uddeholms AB, Uvedsv{\"a}gen, Hagfors, 683 33, V{\"a}rmlands l{\"a}n, Sweden}
}
\email{anton.engman@uddeholm.com}

\renewcommand{\shortauthors}{Chatterjee A, et al.}

\begin{abstract}
There is often a scarcity of training data for machine learning (ML) classification and regression models in industrial production, especially for time-consuming or sparsely run manufacturing processes. Traditionally, a majority of the limited ground-truth data is used for training, while a handful of samples are left for testing. In that case, the number of test samples is inadequate to properly evaluate the robustness of the ML models under test (i.e., the system under test) for classification and regression. Furthermore, the output of these ML models may be inaccurate or even fail if the input data differ from the expected. This is the case for ML models used in the Electroslag Remelting  (ESR) process in the refined steel industry to predict the pressure in a vacuum chamber. A vacuum pumping event that occurs once a workday generates a few hundred samples in a year of pumping for training and testing. In the absence of adequate training and test samples, this paper first presents a method to generate a fresh set of augmented samples based on vacuum pumping principles. Based on the generated augmented samples, three test scenarios and one test oracle are presented to assess the robustness of an ML model used for production on an industrial scale. Experiments are conducted with real industrial production data obtained from Uddeholms AB steel company. The evaluations indicate that Ensemble and Neural Network are the most robust when trained on augmented data using the proposed testing strategy. The evaluation also demonstrates the proposed method's effectiveness in checking and improving ML algorithms' robustness in such situations. The work improves software testing's state-of-the-art robustness testing in similar settings. Finally, the paper presents an MLOps implementation of the proposed approach for real-time ML model prediction and action on the edge node and automated continuous delivery of ML software from the cloud.
\end{abstract}

%
%

%

\maketitle

\section{Introduction}
Machine learning (ML) software has grown in popularity in recent years due to the availability of massive amounts of training data and application size. When training an ML software, in general, the conventional approach is to divide the data into training and testing for regression or classification. A large portion of the ground truth is often used for training, leaving a fraction of the samples for test, usually 25\% or less. A problem arises when insufficient data is available, which takes practically all the ground truth for training and leaves a few samples for testing using the conventional approach. For example, when an event is performed `N' (N $<$ 10) times a day during the workday, such as a manufacturing/production process, it adds up to roughly 300N different events per year. Delivering stable decisions over time is a critical issue in the industrial application. Therefore, testing the robustness of such a software product is an essential quality assurance check. Robustness is a non-functional quality attribute of ML software. Testing the robustness of ML software under limited data conditions is underexplored in the literature. The question remains whether these limited data are sufficient to render a robust ML model using the conventional training-test split strategy.

An instance where there are insufficient data available is in the refined steel industry, specifically vacuum pumping in the Electroslag Remelting (ESR) process \cite{ESRUddeholm2020, ESRModellingSimReview2018, ESROverview2016}. The process includes furnace cleaning \cite{furnaceCleaningRole2018}, which requires vacuum pumping or extracting oxygen from the furnace. As part of the ESR process, when a vacuum pumping event occurs, the sensors attached to the chamber collect data on the total pressure inside the chamber. The pumping process is terminated when the desired pressure threshold is reached. Predicting the minimum pressure of a vacuum chamber early in the pumping process can potentially identify issues such as pump failures and leaks. However, a pumping event occurs once a day on average and generates a few hundred events in a year to split between training and test. In this case, for an ML model, there are only a few hundred occurrences to divide between training and test data. It becomes difficult to obtain the appropriate test data to determine whether the ML model for pressure prediction is overfitting \cite{Overfitting2019}.

Theoretically, to avoid overfitting, one of the following steps is taken \cite{OverfittingTheory2019,CrossValidation2018,Regularisation2007}: (i) removing features such as random forest, (ii) increasing the amount of data available for training,  (iii) performing cross-validation, or with (iv) regularisation, which includes calibrating the hyperparameters for the given data. Data augmentation \cite{augReview2021,augChapter2018} is one such technique for generating slightly altered copies of existing data or creating new synthetic data and boosting the number of training samples available. We use data augmentation in a new industrial application in the ESR process. The data augmentation technique and the testing approach discussed in this paper are based on the general principles of vacuum engineering. Vacuum pumping is used across multiple domains, including fusion power plants \cite{vacPumpFusion2014} and EUV lithography for global silicon manufacturing \cite{euvLith2018}. This paper focused on ESR due to the availability of pumping data from the Uddeholms AB steel company. However, in principle, this augmentation and testing technique can be applied in multiple disciplines where vacuum pumping is part of the manufacturing process.

This study offers a fresh perspective on data augmentation for vacuum pumping from the software testing point of view. The approach presented in this paper first decomposes the data using domain knowledge and then generates random and fresh samples utilizing the concepts of vacuum pumping. The newly generated augmented data are then used to assess ML models' robustness on an industrial scale. Here, careful test scenarios were designed to determine the end-to-end functionality of the ML systems. Each scenario is combined with a test oracle to determine the pass and fail cases during robustness testing. The oracle has been designed from a combination of theory and domain-knowledge background in industrial production. Therefore, ML models are subjected to robustness testing, an important non-functional testing property of ML software \cite{MLTesting}. Here, robustness testing is used to ensure that they work as expected with known data (ground truth or GT), remain within physical constraints, and, in theory, are resilient to future unknown data. To assess the robustness in the presence of limited data, this paper makes the following contributions:

\begin{itemize}
    \item A data augmentation approach for ESR vacuum pumping is developed that uses domain knowledge to generate synthetic data.
    \item A strategy is developed to evaluate the robustness of an ML model based on real and augmented data to compensate for insufficient ground truth test data.
    \item The paper used real data from steel production and ML models on an industrial scale.
    \item An MLOps implementation of the proposed approach.
\end{itemize}

The remainder of the paper is structured as follows. Section \ref{sec:overview} gives an overview of ML training with small samples and vacuum pumping theory to describe the principles underlying the data decomposition and pumping speed dictionary mentioned in the proposed work. The subsequent section \ref{sec:proposed} describes (i) an efficient data augmentation approach for vacuum pumping and (ii) tests to evaluate the robustness of a trained ML model. Section \ref{sec:mlops} shows how the proposed approach is used in industrial production, where training, testing, and real-time prediction and actions of the ML software are carried out automatically, followed by Section \ref{sec:results} which contains experiential results. Section \ref{conclusion} concludes the paper and provides future directions for continuing this study.

\section{Background and related Work} \label{sec:overview}

\subsection{Small sample ML and data augmentation}

Small samples have low statistical power, making it difficult for the ML software to train robustly \cite{smallSample2020}. It is difficult to accurately predict performance from a small set of measured variants, especially if those features interact. An approach to statistical learning performance prediction that considers variability is presented in the paper by Guo et al. \cite{smallSample2013}. The method uses a series of random samples to detect feature interactions step by step without additional effort. The method investigates and exploits the correlation between feature selections and performance to predict performance in a configurable software system. This correlation can be easily revealed by measuring the performance of all software system configurations and then providing direct answers (e.g., which configuration is the fastest). In addition, small sample ML requires a robust validation method. The simulations in Vabalas et al.'s paper \cite{smallSample2019} show that the k-fold cross-validation produces strongly biased performance estimates with small sample sizes, and the bias is still apparent with sample sizes of 1000.

Increasing the training sample size in a principled manner is a constant demand for ML software. Increasing training data with data augmentation improved accuracy in various areas. Two key categories are defined in the survey paper by Wen et al. \cite{augReview2021} on time-series data augmentation: basic and advanced.  Basic approaches include the time domain, the frequency domain, and the time-frequency domain, which include augmentation approaches such as noise injection, flipping, and jittering. Advanced methods include decomposition, statistical generative models, and learning. The survey's primary focus is on class imbalance and data augmentation for members of the minority class. Among other methods, Generative adversarial networks (GANs) have been shown to be able to produce artificial data that are similar to the real data. When faced with the problem of unbalanced data classification, GANs can provide an alternative solution. By generating artificial data that are similar to the original data, and thus augmenting the training dataset, GANs can be used for data augmentation. GANs, for example, in the papers by Shao et al.\cite{GAN2019}, Ortego et al. \cite{GAN2020}, and more recently in Jain et al. \cite{GAN2022} are intended to produce realistic synthesized signals with labels for further use in machine fault diagnosis. A limitation of such methods is that all augmented samples produced may not be physically plausible and may show unrealistic artifacts. A recent study by Lee and Lee \cite{LeeAug2021} presents a unique deep learning methodology for evaluating the interior noise in automobiles. The method uses domain knowledge for data augmentation and for model development, has seen substantial increase in ML software accuracy. Thus, unlike other approaches,, this paper expands on data augmentation in ESR and proposes a new method that combines domain knowledge and ML to test the robustness of such software products.

\subsection{ESR Vacuum Pumping}

\begin{figure}[!htpb]
    \centering
    \includegraphics[scale=0.45]{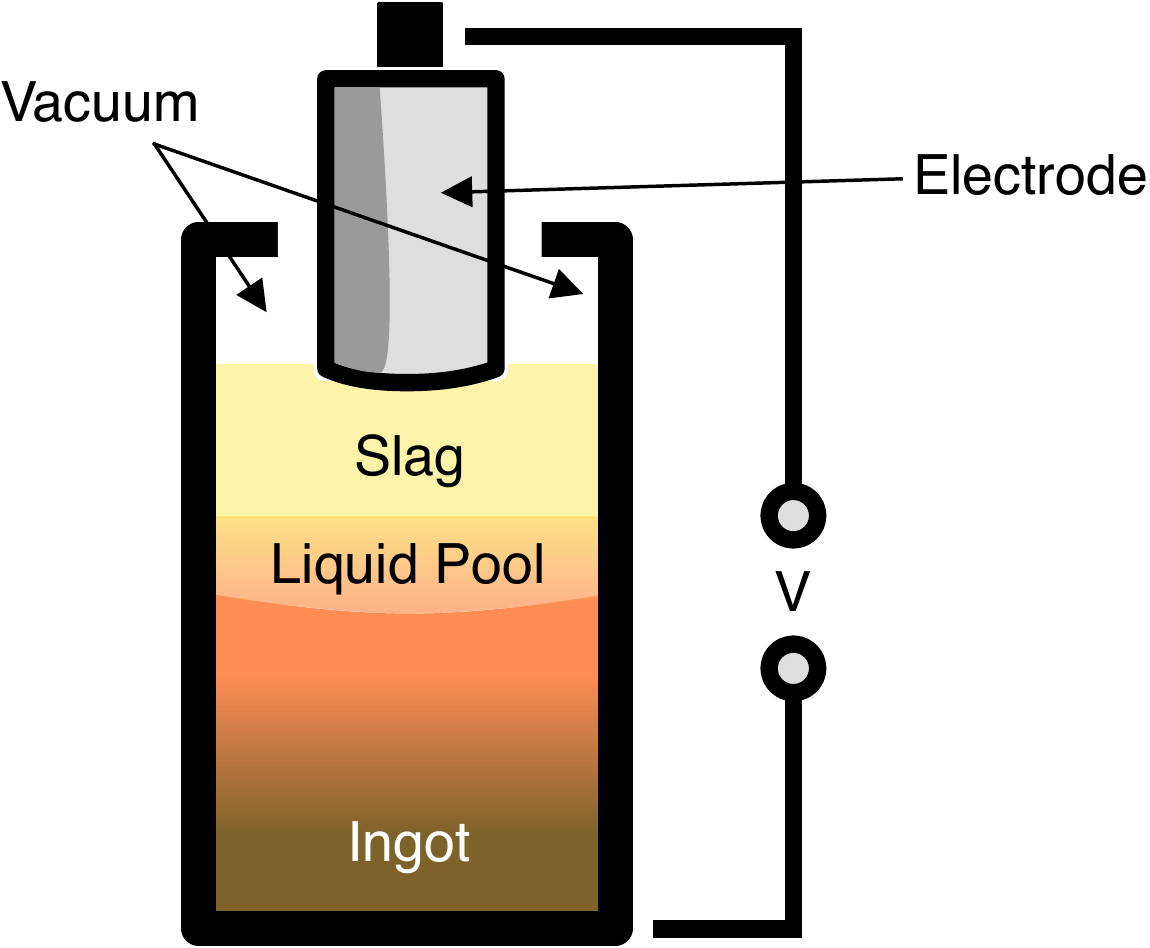}
    \caption{Schematic sketch of ESR.}
    \label{fig:esrDiag}
\end{figure}

The ESR process involves remelting the consumable electrode as the steel is refined and solidified \cite{ESRDiag2017}. Figure \ref{fig:esrDiag} depicts the schematic diagram of the ESR, where the region above the slag requires near-vacuum conditions to avoid contaminants during the process. As a result, the vacuum pumps attached to the ESR pump out excess oxygen or air from the chamber.

Theoretically, vacuum pumps in the ESR process reduce the number of gas molecules in a vacuum system by creating a volume flow determined by the rate at which the pressure in a vacuum system of constant volume decreases over time \cite{CFDVacuum2017}. When pumping occurs, the overall pressure inside a chamber is directly proportional to the amount of gas extracted by the pump, and therefore there is less leakage and flow through the inner surfaces of the chamber. The amount of gas pumped out and the amount of gas that leaks in are constantly changing over time. Dmitrieva et al. \cite{vacuumPumpModel2017} describe the total pressure inside a vacuum system at a time `T' as `$P_T$' with an initial pressure of `$P_0$', the volume of the vacuum pump chamber `$V_c$' the average pumping speed `$S$', the leakage flow `$Q_l$' and the flow from the inner surfaces `$Q_i$', which is:

\begin{equation}
P_T = \frac{Q_l + Q_i}{S} + (P_0 - \frac{Q_l + Q_i}{S}) e^{-\frac{TS}{V_c}}
\label{eq:pumpingEq}
\end{equation}

In the above equation \ref{eq:pumpingEq}, $Q_l$ depends on the total conductance that determines the aggregate bandwidth of all leaks. And, $Q_i$ is proportional to the surface area of the vacuum chamber, a degassing coefficient, and a constant that depends on the material and roughness of the vacuum chamber. However, in practice, a vacuum pump can only lower the pressure to a certain point; and the time taken is referred to as the pump-down time. An approximate model \cite{pumpdownTime2011} with effective pumping speed `$S_T$' at pump-down time `T' is given as:

\begin{equation}
    S_T = \frac{V_c}{T}ln(\frac{P_T}{P_0})
    \label{eq:approx}
\end{equation}

The pumping event is not a common occurrence in real production. For example, in a steel company such as Uddeholms AB in Sweden, for which this study is conducted, the pumping event occurs once a day on average for approximately twenty minutes to pump out oxygen from the vacuum chamber. After pumping for twenty minutes, experts inspect the chamber to ensure that it meets quality standards before moving on to the next phase, such as checking internal pressure. If the quality requirements are not achieved, the chamber is inspected and re-pumped for an additional twenty minutes. The role of ML in this situation is to observe the first few seconds or a minute of pumping and to forecast the minimum pressure. In this case, accurate pressure forecasts save valuable time and increase production efficiency. However, with the limited occurrence of the event, a few hundred samples can be generated in a year of pumping for training and testing that will affect the robustness of the implemented ML model.

\section{The Proposed Approach} \label{sec:proposed}

This paper presents two concepts for the ESR vacuum pumping application with limited training data: (i) data decomposition and augmentation to generate new vacuum pumping data, and (ii) tests for evaluating the robustness of a trained model using the generated augmented data and new test scenarios. 

\subsection{Augmented Sample Generation}
A three-step strategy is designed to generate augmented data and address the scarcity of training data for vacuum pumping. The first step is to decompose the ground truth and interpret the initial pressure and pump-down time. Second, a dictionary of pumping speeds is constructed that is independent of the initial pressure. The final step generates sparse representations for the dictionary and produces a fresh set of augmented samples. The following subsections address these steps. 

\subsubsection{Decomposing the ground truth data} \label{prop:mle}
Each pumping event is a separate and distinct occurrence. For a given vacuum chamber (i.e., a constant volume), equation \ref{eq:approx} mandates that one must understand `$S_T$', `$T$', and the `$P_0$' in order to interpret `$P_T$'. At minimum pressure, all three mentioned variables are scalars. However, a vector is formed for $S_T$ when we look at the entire pumping period (t = 0 to T). We take the approach to decompose the ground-truth by interpreting the $P_0$ and $T$ distributions and extracting the $S_T$ into a dictionary or a matrix containing only the unique $S_T$ vectors. As a result, we use the central limit theorem and the law of large numbers \cite{CLT2018} to approximate the initial pressure and pump-down time as a Gaussian distribution. Although future pumping occurrences are unknown, this assumption will, in principle, assist a regression/pressure prediction model in learning about the different pumping events in a proactive manner. The mean and standard deviation of the distributions for $P_0$ and $T$ are computed using the maximum likelihood estimation from ground-truth data.

\subsubsection{Extracting a dictionary of pumping speeds} \label{prop:dict}
After obtaining the $P_0$ and $T$ distributions, the next step is to obtain the pumping speeds $S$ per unit volume at regular intervals. Because the amount of gas coming in and going out changes over time, we extracted all $S$ from Equation \ref{eq:approx}. The collected $S$ was then reduced to the independent pumping speeds to a dictionary $D_S \in \R^{X \times Y}$ ($D_S \subset S$) assuming a simplex and linear mixing. Mathematically, for an acceptable residual error $\epsilon$ and a representation $\psi$, it is given as:

\begin{equation}
    \min_{D_S \in \R}{||D_S||_{row, 0}} \text{ s.t. } ||S - D_S \psi||_2 \leq \epsilon
\end{equation}

\subsubsection{Generating augmented samples}

Once the parameters $P_0$ and $T$ from section \ref{prop:mle}, and a dictionary $D_S$ mentioned in section \ref{prop:dict} for the vacuum chamber and pump setup are determined, the next step is to generate as many augmented samples M ($M \gg N$) as are required. For each of the $M$ samples, a sparse vector $\psi$ ($\psi \in \R^{Y \times 1}$) is generated at random, resulting in a unique pump speed that is a fractional combination of historical events. The mathematical conditions of $\psi$ are:

\begin{equation}
    \forall \psi_y \subset \psi \text{, } \psi_y \geq 0 \text{ and } ||\psi||_1 = 1
\end{equation}

\begin{figure}[!htpb]
    \centering
    \includegraphics[scale=0.7]{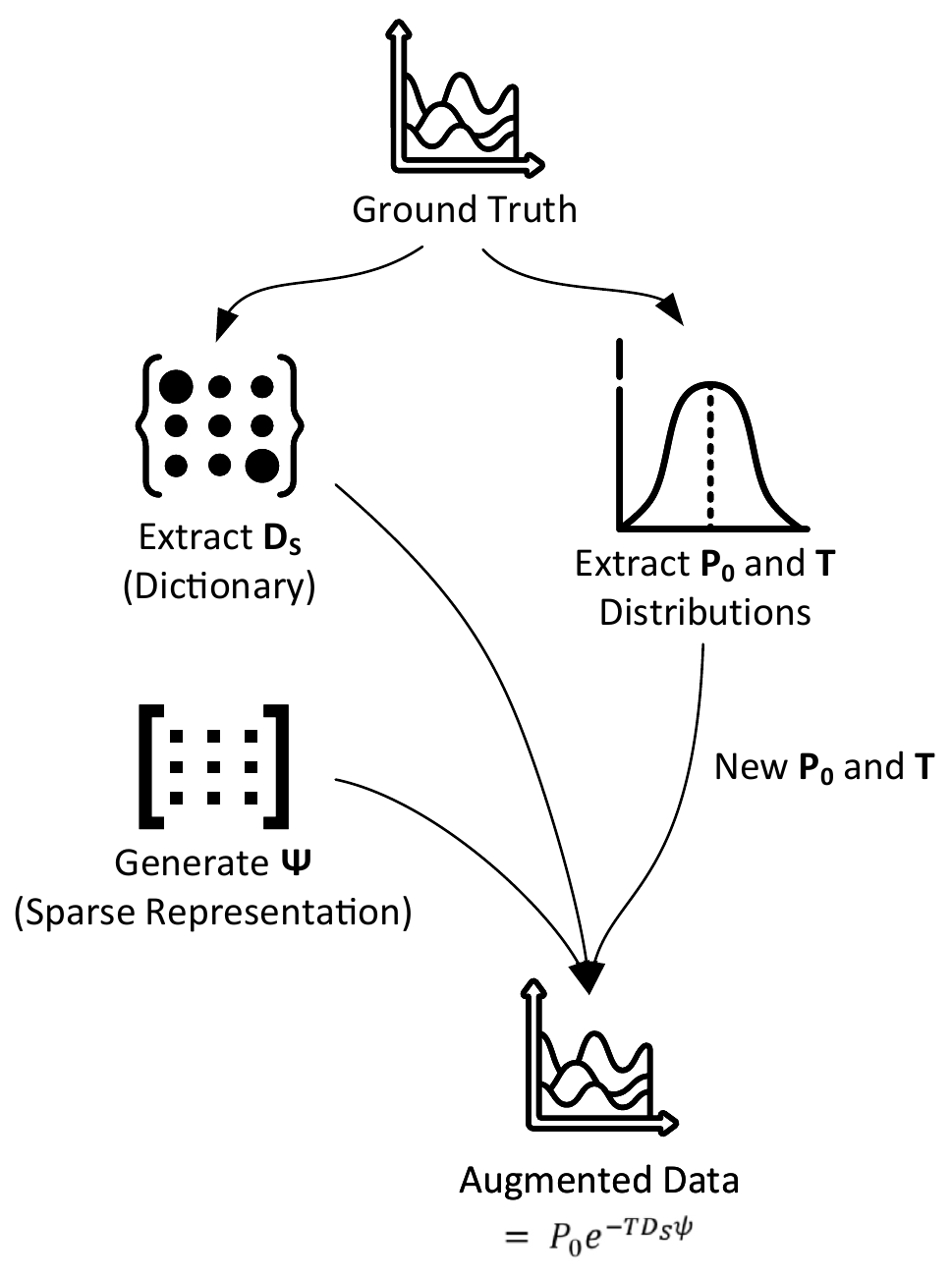}
    \caption{A diagram showing the proposed data augmentation technique for ESR vacuum pumping.}
    \label{fig:augFlow}
\end{figure}

After that, a random initial pump speed $P_0$ and pump-down time $T$ are generated for their respective distributions, and a convolution operation with equation \ref{eq:approx} creates a unique augmented sample. This augmented sample differs from the ground truth data while remaining within the constraints of vacuum pumping's physical capabilities. The procedure is repeated M times to yield M unique samples. A diagram of the proposed data augmentation is depicted in Figure \ref{fig:augFlow}.

\subsection{Tests to Evaluate Robustness of a Trained ML Model}
Once an ML model is trained using augmented data, it must be tested to ensure that it is robust. We propose three test scenarios to evaluate the robustness of the model. Each of the three test scenarios addresses a unique concern, which are as follows:

\begin{itemize}
    \item \textbf{Feasibility test scenario:} is to see if the trained model remains within the physical limits of vacuum pumping.
    \item \textbf{Ground truth testing:} examines the trained model's accuracy and its performance when applied to known data.
    \item \textbf{Volume enclosed testing:} under a given residual error examines how well the trained model should operate under the constraints of vacuum pumping (e.g., in case of a drift in the data or changes in the pumping conditions).
\end{itemize}

Depending on the application's sensitivity, the results from the scenarios are collected into a test oracle that determines if an applied robustness test of the machine learning model is passed or failed.

\subsubsection{Feasibility Test Scenario}
The feasibility test scenario ensures that a trained model stays within the physical limits of vacuum pumping. This scenario warrants that no direction from initial pressure and pump-down time is impractical, i.e., the predicted $P_T$ is always positive when $P_0$ and $T$ are positive. Mathematically, it is:

\begin{equation}
    \forall P_0 > 0 \text{ and } T > 0, P_T > 0
\end{equation}

\subsubsection{Testing the ground truth}
After the feasibility test, we have a scenario that uses the ground-truth and a scenario that uses augmented data. The second test scenario evaluates the residual error of the data from the ground-truth test data. The mean absolute error (MAE, equation \ref{eq:mae}), also known as the Manhattan distance or the $\ell_1$ norm of the residual error divided by the number of observations, and the goodness-of-fit measure: R-squared ($R^2$, Equation \ref{eq:r2}) value. An additional $\ell_\infty$ with the real and augmented data shows the maximum expected prediction error for the given data.

\begin{equation}
    MAE(a, b) = \frac{\sum_{i = 1}^{n} | a_i - b_i |}{n}
    \label{eq:mae}
\end{equation}

\begin{equation}
    R^2(a, b) = 1 - \frac{\sum_{i}^n (a_i - b_i)^2}{\sum_{i} (a_i - \bar{a})^2}, \text{ where } \bar{a} = \frac{1}{n}\sum_{i=1}^n a_i
    \label{eq:r2}
\end{equation}

\subsubsection{Volume enclosed under a given residual error}
\begin{figure}[!htpb]
    \centering
    \includegraphics[scale=0.42]{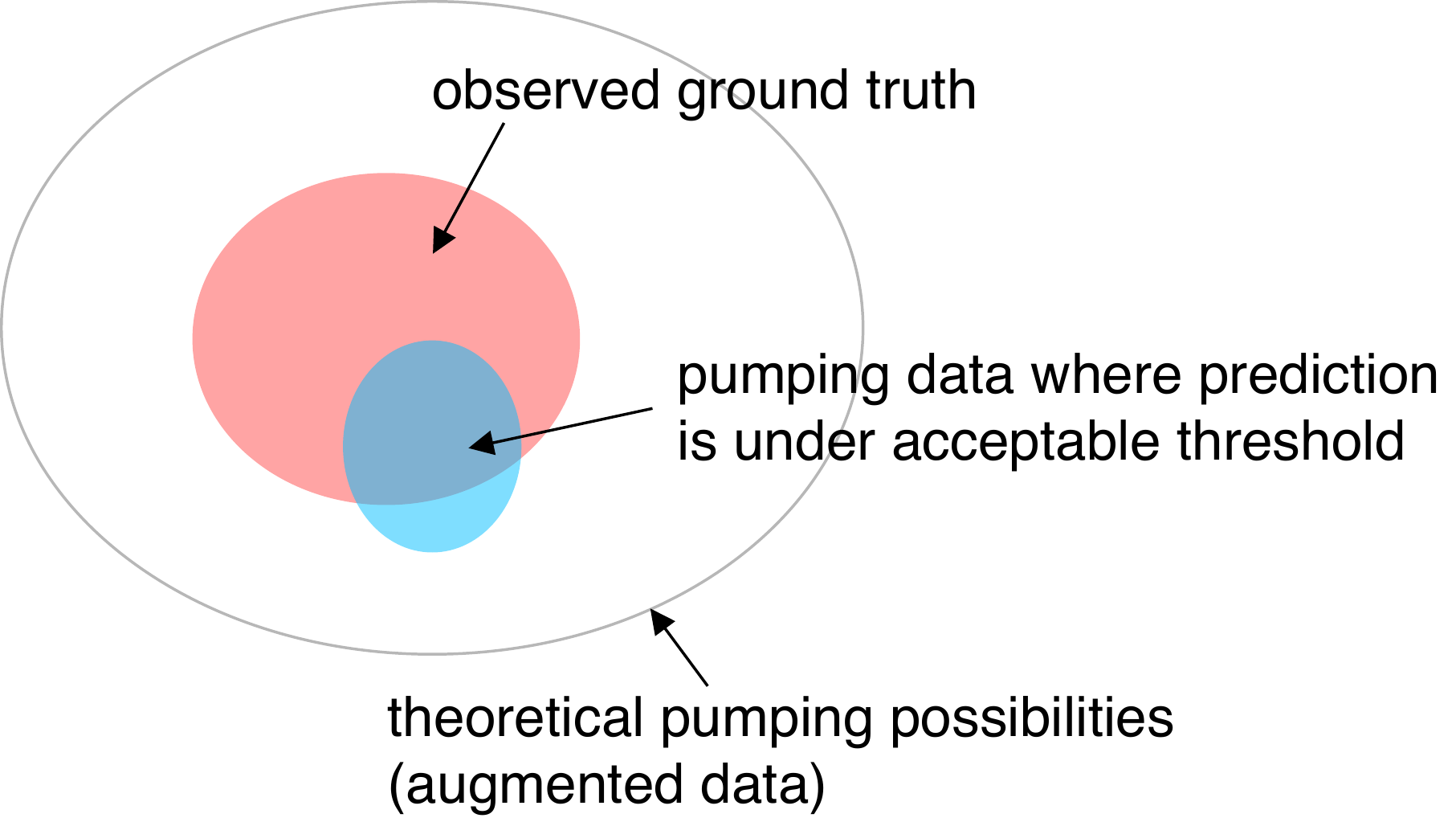}
    \caption{A two-dimensional illustration of the area enclosed by the residual error (under a desired threshold) of pumping measurements.}
    \label{fig:volTheory}
\end{figure}

The third test scenario involves estimating the volume of occupied space within a specified residual error threshold. A higher volume indicates a more robust method for the given data and threshold. A graphical illustration of the concept is shown in Figure \ref{fig:volTheory}.

This test retrieves augmented samples whose residual error is less than a specified threshold, `t'. Afterward, the linearly independent columns are located. In the end, the determinant method \cite{volSimplex2021} is used to compute the volume, which is:

\begin{equation}
    |V(P)| = \frac{\sqrt{\text{det}(\bar{P}^T\bar{P})}}{(d-1)!}
\end{equation}
\begin{equation*}
    \text{Where, } \bar{P} = [P_1 - P_d, P_2 - P_d,\text{ ... , }P_{d-1} - P_d]
\end{equation*}

\subsubsection{Test Oracle}

\begin{figure}[!htpb]
    \centering
    \includegraphics[scale=0.8]{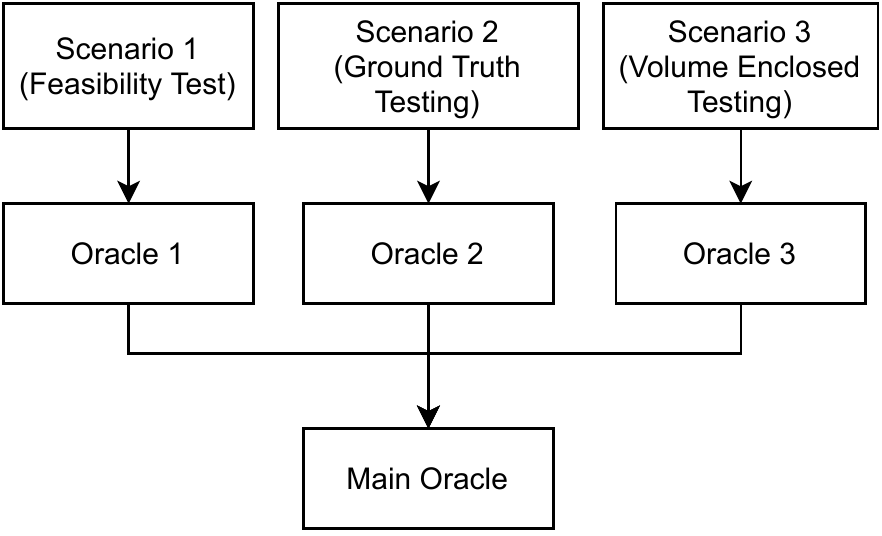}
    \caption{Robustness testing test oracle for the ESR vacuum pumping}
    \label{fig:oracleDiag}
\end{figure}

In the final phase, the results/output of the three test scenarios are fed into three sub-oracles and then to the main oracle. Figure \ref{fig:oracleDiag} displays the flow diagram of the sub and main oracles.

\begin{itemize}
\item \textbf{Oracle 1}: The first sub-oracle (or oracle 1) is to identify whether a vacuum pump pressure prediction ML model predicts positive values with the augmented samples. If any of the augmented samples as input returns a negative pressure, the ML model under test is considered a failure of the test. Mathematically, it is given by:
    \begin{align}
        \sum P_T > 0 = \left \{ \begin{array}{ccc}
        1 & , & \text{Pass}\\
        0 & , & \text{Fail}
        \end{array}
        \right.
    \end{align}
    
\item \textbf{Oracle 2}: The second sub-oracle gets its input from scenario 2 (or ground-truth testing). The sub-oracle compares the predicted values/scores against acceptable thresholds determined by field experts. Selecting thresholds is about determining how pure each block of steel should remain. In principle, a higher minimum pressure means that more gas molecules remain in the chamber, reacting with the steel during subsequent production processes after vacuum pumping. The output of this sub-oracle determines whether a ML model meets the threshold requirements of the application. For example, for a given threshold of `$t$' for the MAE, the pass-fail criteria is:
    \begin{align}
        MAE \left \{ \begin{array}{ccc}
        \ge t & , & \text{Pass}\\
        < t & , & \text{Fail}
        \end{array}
        \right.
    \end{align}
    
\item \textbf{Oracle 3}: The third sub-oracle inputs two volume measurements from the volume enclosed testing scenario: the volume of the augmented sample space in its entirety ($V_{tot}$) and the volume of the augmented data space ($V_{t}$), which is under the field expert determined threshold. The ratio of the two volumes is calculated and the pass-fail criteria is given by a threshold `$t_v$' as:
    \begin{align}
        \frac{V_{t}}{V_{tot}} \left \{ \begin{array}{ccc}
        \ge t_v & , & \text{Pass}\\
        < t_v & , & \text{Fail}
        \end{array}
        \right.
    \end{align}
Once an ML model under test passes the sub-oracle, the $V_t$ value is passed on to the main oracle.
    
\item \textbf{Main Oracle}: The main oracle takes input from all three sub-oracles and determines the robustness of the ML model. A robust model must stay within the physical boundary (i.e., passed the oracle 1), meet the business requirements (i.e., passed the oracle 2), and be able to adapt to slight changes in the ground truth data (i.e., passed oracle 3). However, suppose that an ML model under test passes oracles 1 and 2 only. In that case, it suggests that the ML model fairs well with the ground truth data, but any deviation or drift may produce unreliable predictions or be less robust. For this reason, the main oracle considers an ML model under test a robust model when it has passed all three sub-oracles. Finally, among several ML models under test that passed the main oracle; the most robust ML model is the one with the largest volume, demonstrating resilience towards changes in the ground truth while delivering the required accuracy in the current data.
\end{itemize}

\section{MLOps Implementation} \label{sec:mlops}

\begin{figure*}[!htpb]
    \centering
    \includegraphics[scale=0.65]{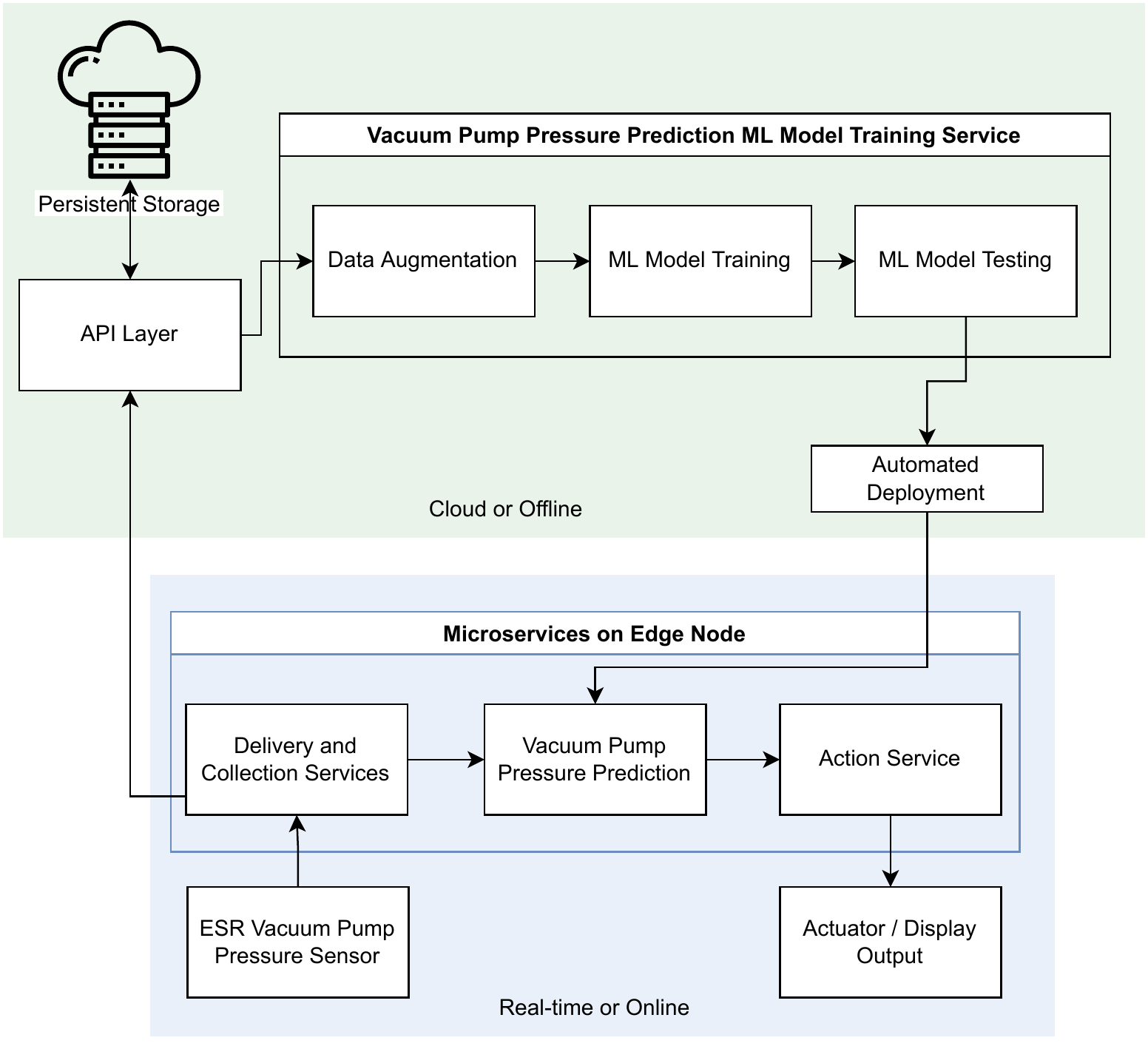}
    \caption{Data pipeline for MLOps implementation of the proposed approach in industrial setting enabled by a combination of (i) sensor, actuator, and Edge microservices (light blue) for real-time analysis/prediction and action, and (ii) Cloud services (light green) for offline ML software training and testing.}
    \label{fig:overviewFlow}
\end{figure*}

Continuous and automated ML software development, also known as MLOps, is becoming increasingly popular as a result of the success of continuous software development, DevOps in particular. This paper describes an early-stage implementation to continuously deliver ML software in the ESR for the proposed data augmentation and testing approach. The architecture outlined in figure \ref{fig:overviewFlow} shows the proposed implementation with services and microservices split between cloud and edge node. Services like ML software training and retraining and the persistent storing of pumping data are executed in the cloud. Microservices at the edge operate in real-time, making decisions and taking action automatically.

\subsection{Real-time prediction and action}
Automated and real-time decisions in production are made possible by containerized microservices on the Edge node. A great tool for creating containerized applications is Docker\footnote{https://www.docker.com/} \cite{docker2018}. During vacuum pumping, the `delivery and collection' microservice receives real-time data from pressure sensors attached to the ESR. The real-time streaming data pipelines and applications built with Apache Kafka\footnote{https://kafka.apache.org/} is an example of `delivery and collection'. After `T' seconds after pumping, delivery and collection microservice transmits the collected data to the `vacuum pump pressure prediction' microservice, where the ML model trained by our proposed approach is run. The ML model predicts whether or not the desired minimum pressure will be reached and sends the predicted minimum value to the next microservice – `action'. The `action' microservice signals the actuator to stop pumping when the predicted minimum does not meet the expected requirements in a fully autonomous setup. Alternatively, or in addition, the `action' microservice sends a signal to display the predicted minimum for a human field operator to take appropriate steps. The `action' microservice is customizable based on the business needs.

\subsection{Continuous and automated ML software training and deployment}
In order to provide continuous delivery of the ML vacuum pump prediction software, data flows from the sensors attached to ESR to the "delivery and collection" microservice on the edge. The pumping data is subsequently forwarded to the cloud for long-term `persistent' storage. The API layer in the cloud offers an interface between microservices on the edge, cloud services, and persistent storage. The ML software with the proposed approach is then trained with the historical pumping data, and when an ML model passes our proposed main oracle, a new version of the ML prediction software container is automatically deployed to the Edge. At the current stage, the ML model is trained periodically after a few months; however, future work in the architecture will involve the research on additional microservices to detect and classify model degradation and trigger a re-training automatically.

\section{Experimental Evaluation and Results} \label{sec:results}

\subsection{Dataset and ML Methods}
\begin{figure*}[!htpb]
    \centering
    \subfloat[ESR M]{\includegraphics[scale=0.33]{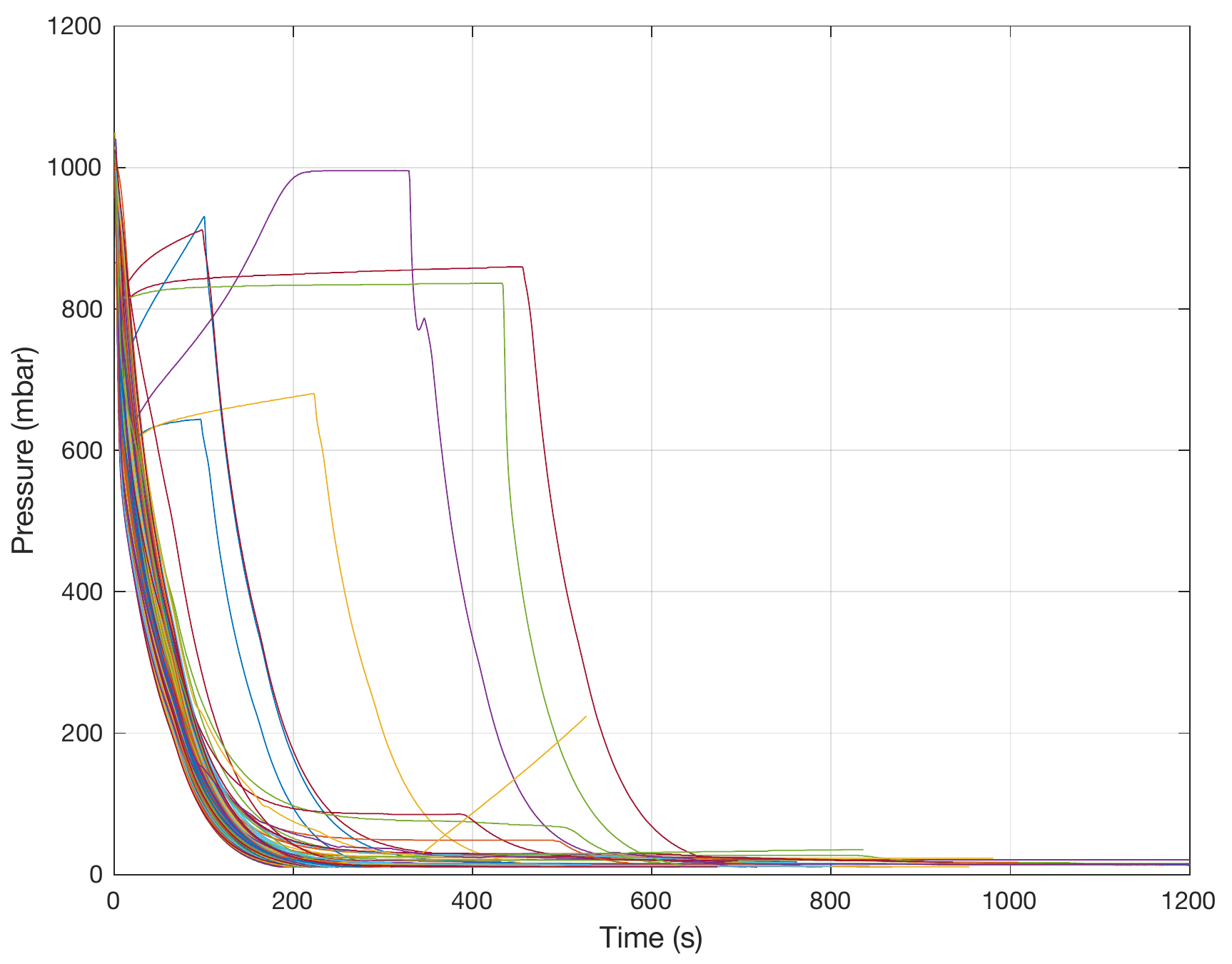}}
    \subfloat[First minute of (a) taken as input for prediction]{\includegraphics[scale=0.33]{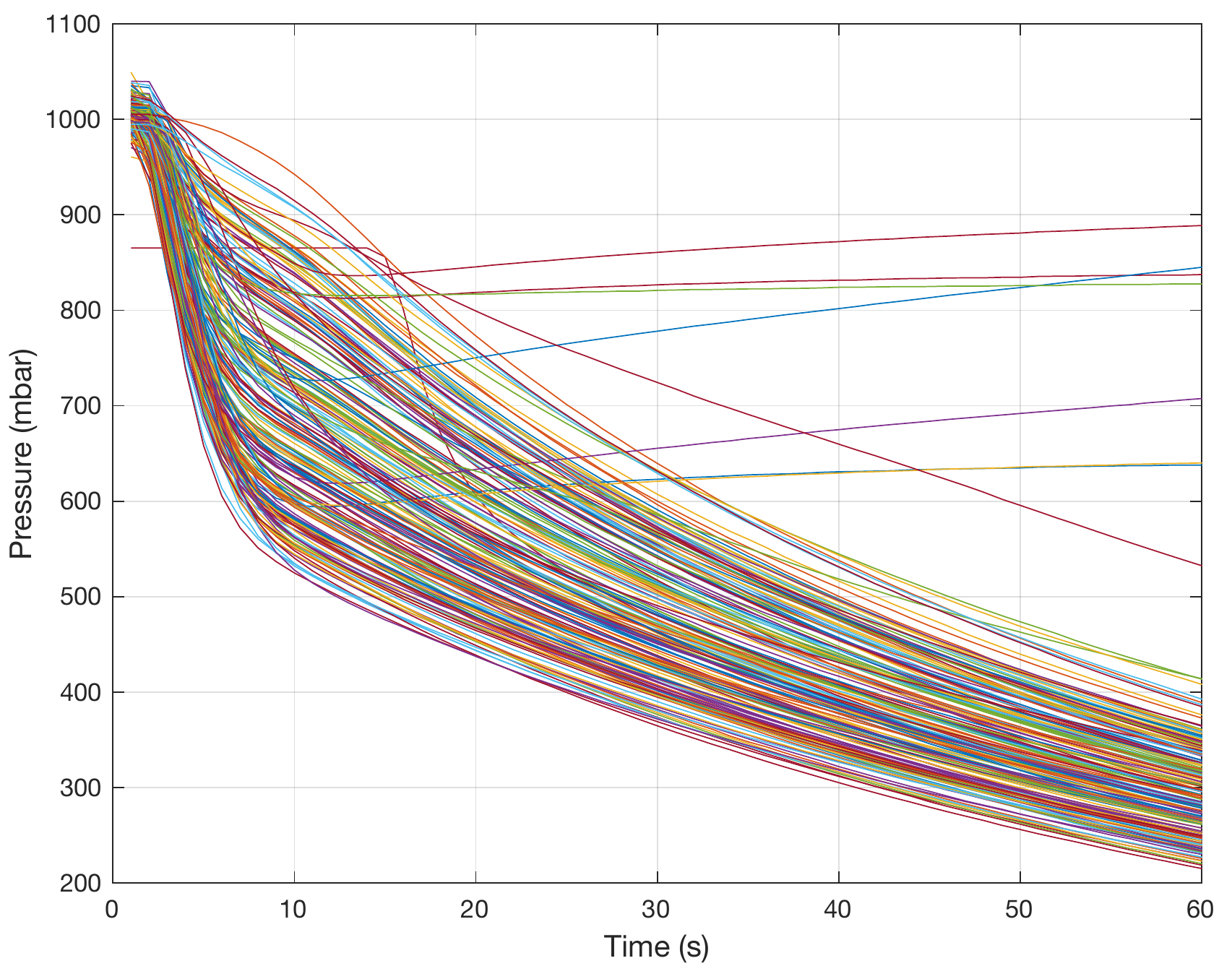}}
    \subfloat[ESR S]{\includegraphics[scale=0.33]{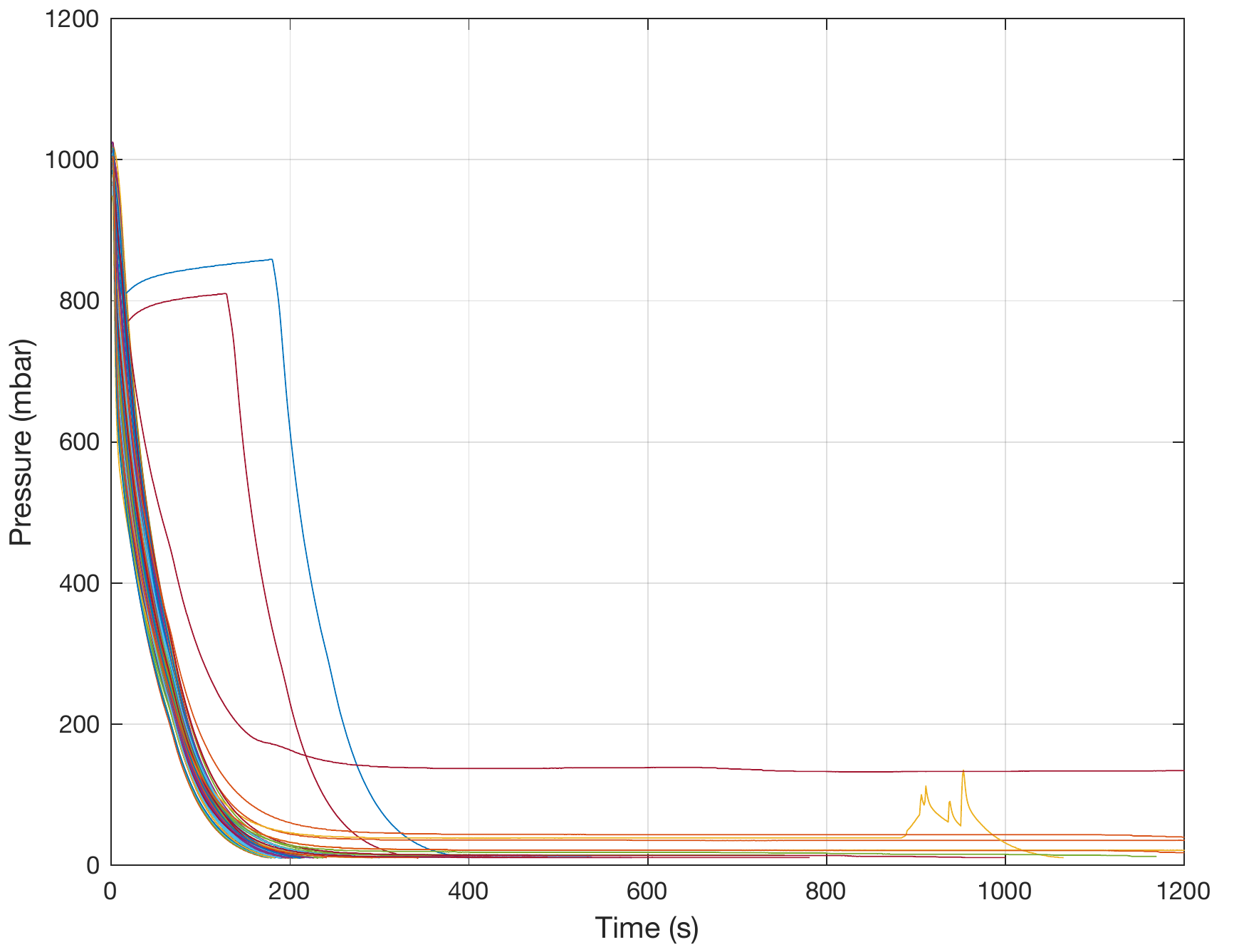}}
    \caption{Two ESR pumps are depicted in this figure: pump M in (a) and pump S in (c). Subfigure (b) displays pumping data from the first minute that is utilized as input to a ML model for predicting the minimum pressure.}
    \label{fig:ESR_GT}
\end{figure*}

The experiments carried out in this paper are using proprietary data from Uddeholms AB on two furnaces of equal volume and manufacturing specifications, termed `furnace M' and `furnace S.' From January to October 2021, Furnace M has 203 events (subfigure (a) in figure \ref{fig:ESR_GT}), and Furnace S has 107 events (subfigure (c) in figure \ref{fig:ESR_GT}). For this experiment, the goal of the pressure prediction model is to predict the minimum pressure (in mbar) reached after twenty minutes of pumping based on the data collected during the first minute of pumping (subfigure (b) in Figure \ref{fig:ESR_GT}).

Three machine learning approaches are evaluated in this paper using the default settings in MATLAB's: (i) Optimizable Ensemble\footnote{https://www.mathworks.com/help/stats/fitensemble.html} (Opt Ensemble), (ii) a single-layered Neural Network (NN)\footnote{https://www.mathworks.com/help/stats/fitrnet.html}, and (iii) Optimizable Gaussian Process Regression\footnote{https://www.mathworks.com/help/stats/fitrgp.html} (or Opt GPR). Bayesian optimization\footnote{https://www.mathworks.com/help/stats/hyperparameter-optimization-in-regression-learner-app.html} was used to optimize the models' hyperparameters during run-time. Similar optimizers exist in other platforms, such as Hyperopt in Python \cite{hyperopt2019} and the paper by Bergstra et al. \cite{hyperopt2011}, which use random search and two new greedy sequential methods based on the expected improvement criterion to optimize hyper-parameters. The hyperparameters are mentioned in table \ref{tab:hyp}.

\begin{table}[!htpb]
\centering
\caption{Table showing the optimized Ensemble and GPR hyperparameters and the default hyperparameters of the single layered NN.}
\begin{tabular}{|c|c|} \hline
\textbf{Model} & \textbf{Hyperparameters}\\ \hline
& Method: Bootstrap aggregation (bagging)\\
& Number of learners: 496\\
Opt Ensemble & Minimum leaf size: 5\\
& Number of predictors to sample: 59\\ \hline
& Layer size: 10\\
NN & Activation: ReLU\\
& Standardize data: Yes\\ \hline
& Sigma: 7.2e-02\\
& Basis: Zero (empty matrix) \\
Opt GPR & Kernel function: Nonisotropic Exponential\\
& Standardize data: Yes\\\hline
\end{tabular}
\label{tab:hyp}
\end{table}

Our proposed data augmentation method and the classical/traditional approaches under test are both trained on like-for-like conditions with MATLAB's Bayesian hyperparameter optimizer in this paper. Using the conventional 80\%-20\% training-test split approach (referred to as `classic' in tables and figures) and 100,000 augmented samples (referred to as `aug' in tables and figures), all three approaches were trained using furnace M data. Classic methods use 80\% of GT data for training, while aug approaches train with the augmented samples. After training, the models were tested on data from furnace S.

\subsection{Data decomposition and augmentation}

\begin{figure*}[!htpb]
    \centering
    \subfloat[Initial pressure ($P_0$)]{\includegraphics[scale=0.33]{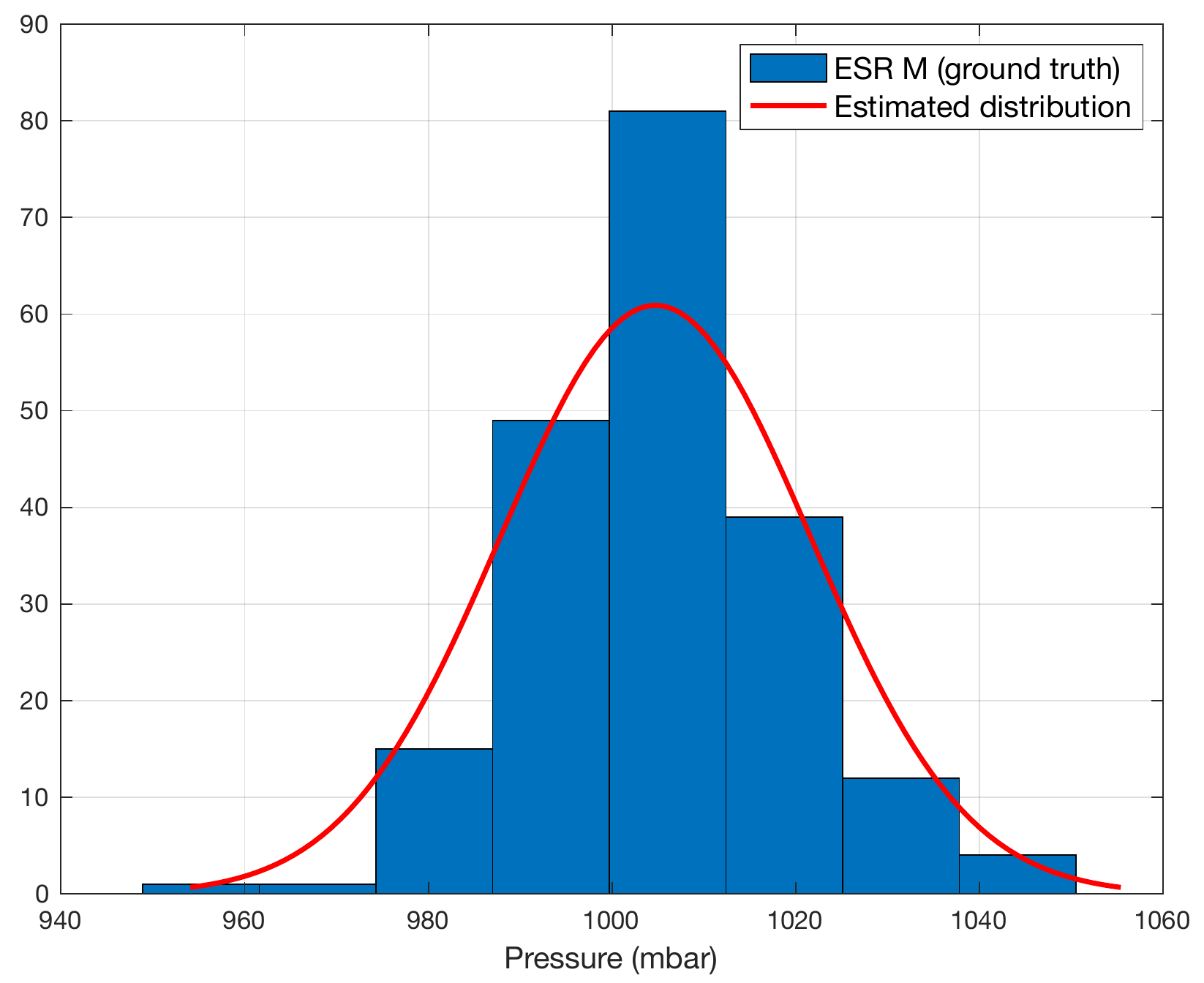}}
    \subfloat[Pump-down time (T)]{\includegraphics[scale=0.33]{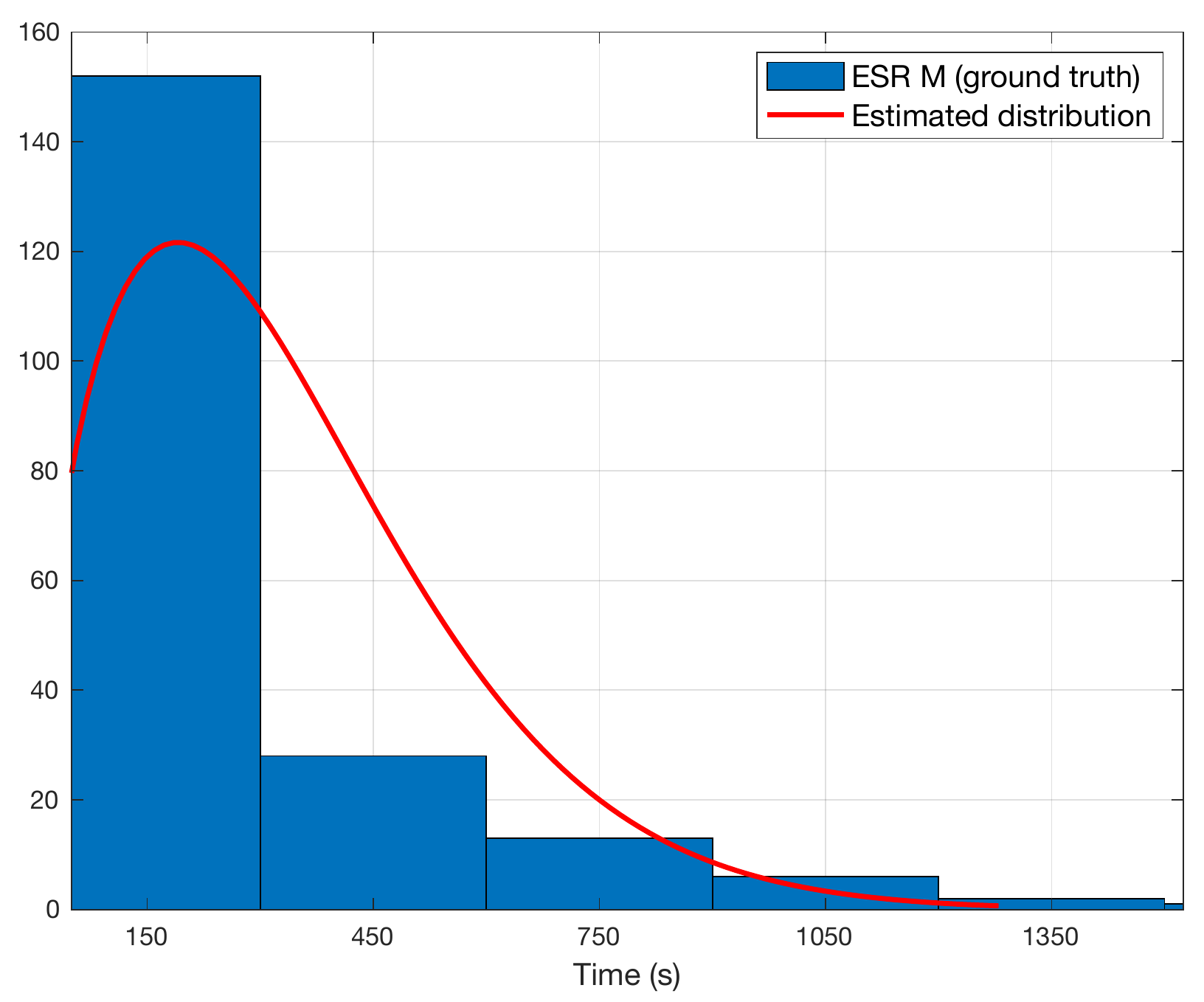}}
    \subfloat[Pumping speed dictionary]{\includegraphics[scale=0.33]{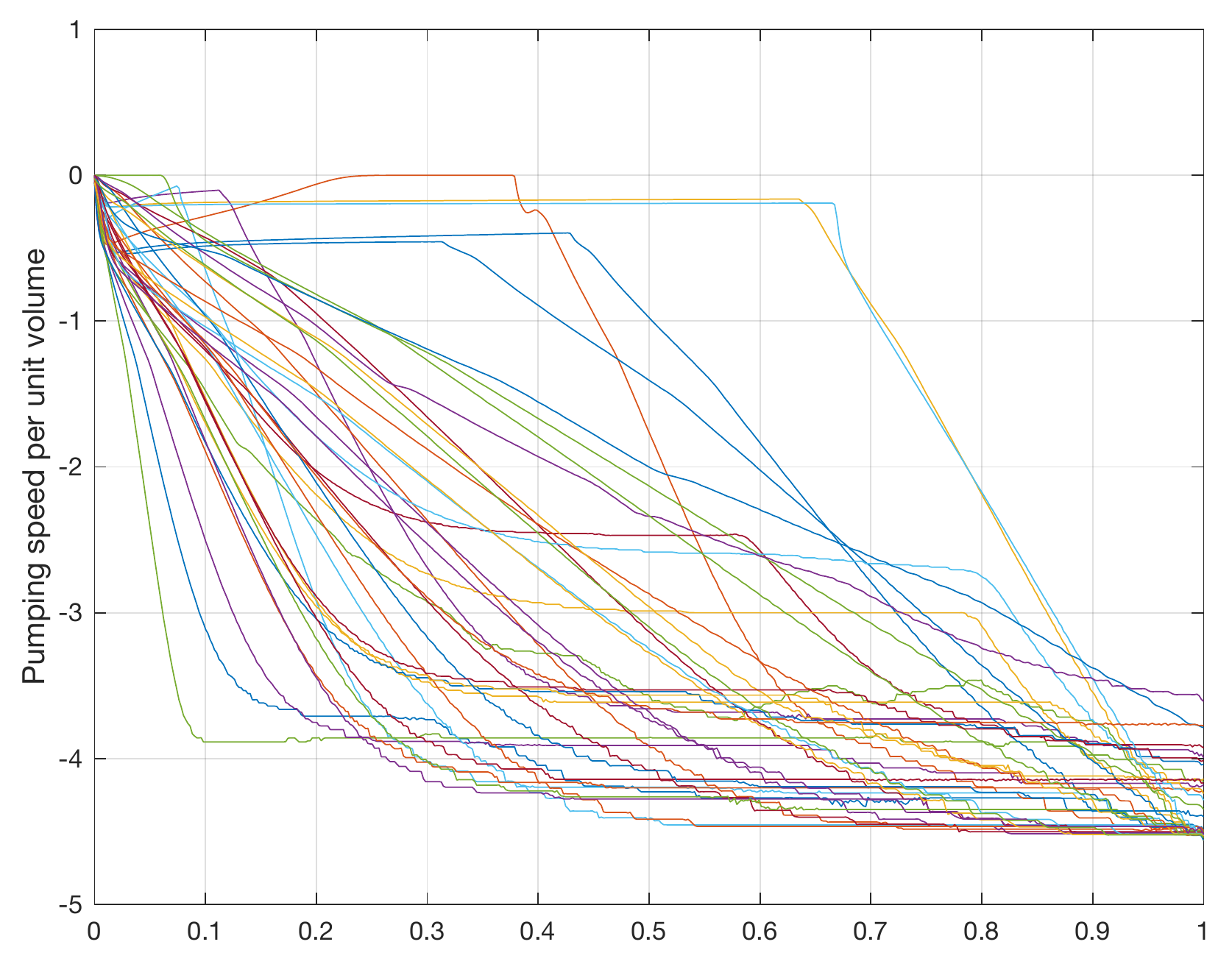}}
    \caption{Figure showing the distributions of the ESR M's initial pressure in (a) with a mean and standard deviation 1e+03 $\pm$ 16.84, and pump-down time in (b) with a mean and standard deviation of 333.59 $\pm$ 262.52. Subfigure (c) shows the extracted pumping speed dictionary.}
    \label{fig:ESRM_decompoition}
\end{figure*}

The first stage in data augmentation of the pump M data is using maximum likelihood to estimate the distributions of initial pressure ($P_0$) and pump-down time ($T$). The distributions of $P_0$ and $T$, respectively, are depicted in subfigures (a) and (b) in figure \ref{fig:ESRM_decompoition}.  The pumping speeds ($S_T$) are then calculated using equation \ref{eq:approx}. After that, the GT data is converted from temporal domain to a time-independent sparse domain. In other words, all individual pumping data with their own length from t = 0 to the pump-down time (T) are interpolated to the same length using cubic spline interpolation\footnote{https://www.mathworks.com/help/matlab/ref/spline.html}. With an average pump-down time of 333s, we selected a dictionary resolution or length of 500 ($\approx$1.5x) to capture minute changes in pump speeds. The collection of all pumping speeds is then subset to the independent pumping speeds with a greedy dictionary learning approach presented in algorithm \ref{alg:SDictionary}.

\begin{algorithm}[!htpb]
\caption{Pseudocode for the extraction of the pumping speed dictionary}
\label{alg:SDictionary}
\begin{algorithmic}[1]
\State \textbf{Input}: S
\While{\textit{stopping criteria for} $\epsilon$}
\State $\psi \gets f(D_s, S)$ \Comment{Representation}
\State $\epsilon \gets S - D_s \times \psi$ \Comment{Residual error}
\State $S_m \gets \arg\max||\epsilon||_p$ \Comment{$S_m$: element with max $\epsilon$}
\State $D_s \gets D_s \bigcup S_m$ \Comment{Add selected $S_m$ to $D_s$}
\EndWhile
\State \textbf{Output:} $D_s$
\end{algorithmic}
\end{algorithm}

\begin{figure*}[!htpb]
    \centering
    \includegraphics[scale=0.4]{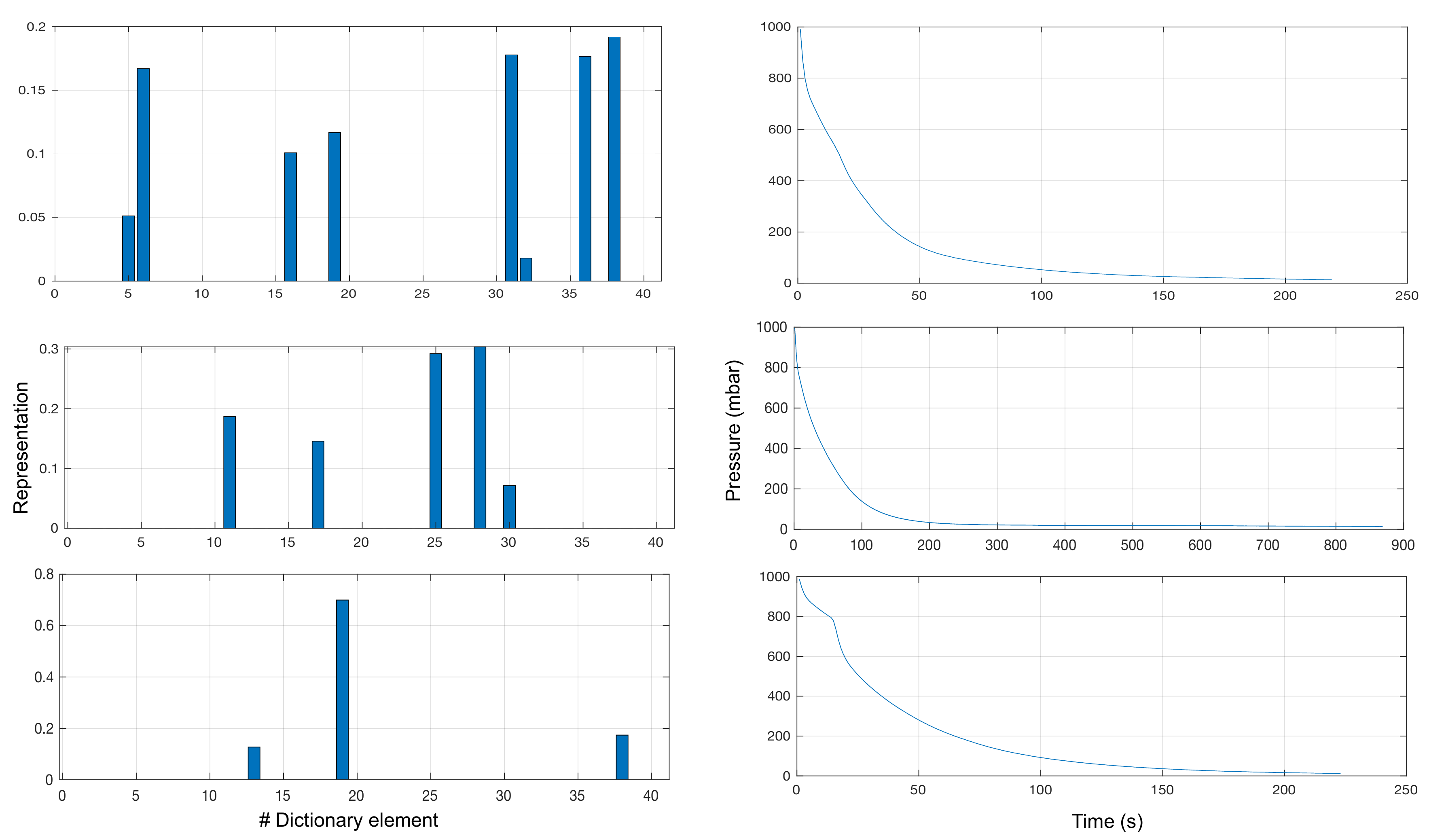}
    \caption{Three examples of augmented samples, each with a pump-down time of 218, 869, and 222 seconds, respectively (top to bottom) and with initial pressures of 1007.9, 1012.3, and 996.8 mbar respectively.}
    \label{fig:exPumpingData}
\end{figure*}

In principle, the representation is an $L_0$ minimization problem which is NP-hard\cite{L0min}, is often approximated to $L_1$ or in this case, we choose the greedy approach, which is $D_s^T \times S$. This approach iteratively selects the pump speed with the highest residual that is orthogonal to the previously selected pumping speeds. With a small residual error of 1e-03 for $\epsilon$, the dictionary selects 40 different pumping speeds. The output dictionary is shown in figure \ref{fig:ESRM_decompoition}c.

After decomposing ESR M data, a random sparse representation and a random initial pressure ($P_0$) and pump-down time ($T$) are generated for augmented data generation. $P_0$ and $T$ are generated with Pearson system random numbers\footnote{https://www.mathworks.com/help/stats/pearsrnd.html} within the limits of the GT data. Figure \ref{fig:exPumpingData} shows three examples of augmented samples.

\begin{figure}[!htpb]
    \centering
    \subfloat[100k samples of augmented data]{\includegraphics[scale=0.45]{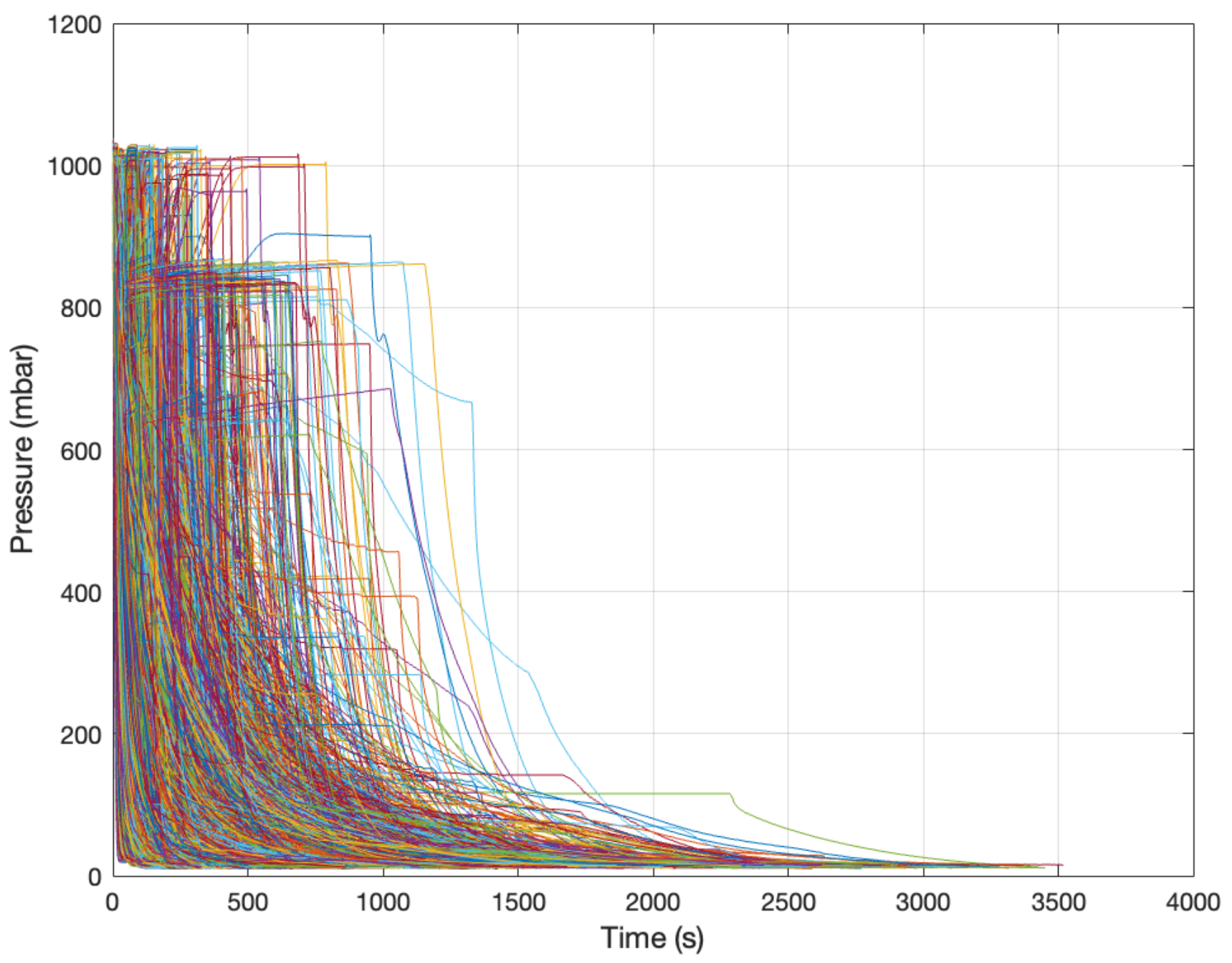}}\\
    \subfloat[GT and augmented data differences]{\includegraphics[scale=0.45]{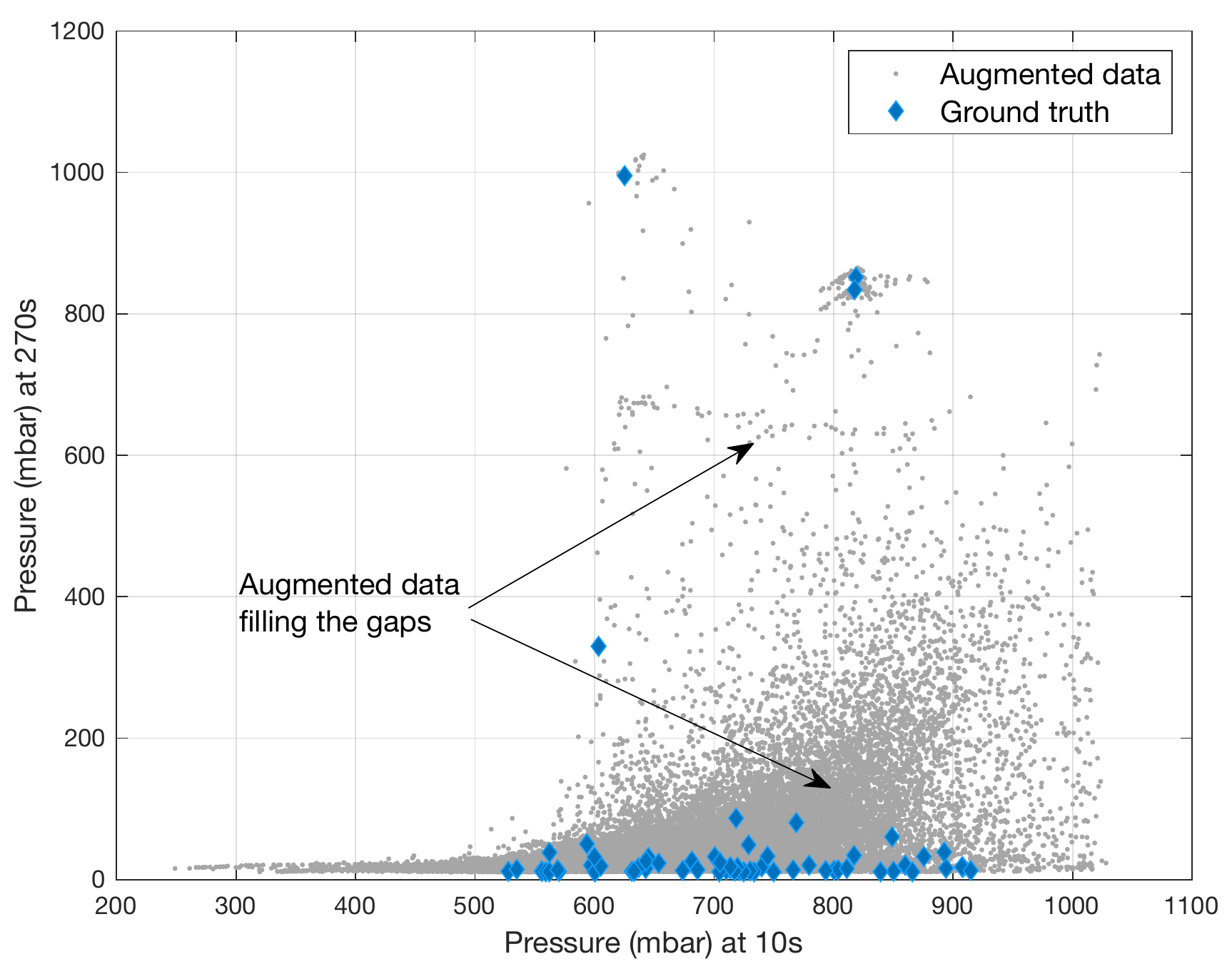}}
    \caption{Augmented pumping data from decomposed ESR M GT.}
    \label{fig:ESRM_augmentation}
\end{figure}

The procedure is repeated for 100k augmented samples in this experiment. The augmented data of all samples is shown in Figure \ref{fig:ESRM_augmentation}.

\subsection{Testing the ML models}

\begin{table*}[!htpb]
\centering
\caption{Residual error of the six models in furnace M, furnace S, and the maximum error observed in the augmented data. Since aug models were trained using augmented samples, they use 100\% of GT data for testing, unlike classic approaches that use only 20\% of GT data for testing.}
\begin{tabular}{|c|ccc|ccc|c|c|} \hline
 & \multicolumn{3}{c|}{\textbf{furnace M}} & \multicolumn{3}{c|}{\textbf{furnace S}} & \textbf{aug} & \textbf{MAE} \\
\textbf{Model} & MAE & $R^2$ & $\ell_\infty$ & MAE & $R^2$ & $\ell_\infty$ & $\ell_\infty$ & $|$S-M$|$ \\ \hline
Opt Ensemble (classic) & 1.03 & 0.95 & 14.76 & 1.38 & 0.94 & 22.40 & 9.04 & 0.35 \\
Opt Ensemble (aug) & 1.20 & 0.93 & 13.56 & 1.00 & 0.98 & 21.02 & 6.37 & 0.20 \\ \hline
NN (classic) & 1.72 & 0.48 & 2.34 & 2.08 & 0.47 & 19.69 & 124.54 & 0.36 \\
NN (aug) & 1.45 & 0.80 & 13.21 & 1.29 & 0.91 & 21.11 & 7.63 & 0.16 \\ \hline
Opt GPR (classic) & 1.21 & 0.58 & 8.80 & 1.28 & 0.83 & 25.29 & 15.39 & 0.07 \\
Opt GPR (aug) & 1.23 & 0.75 & 14.17 & 1.16 & 0.93 & 22.12 & 8.04 & 0.07 \\ \hline
\end{tabular}
\label{tab:test2}
\end{table*}

\subsubsection{Feasibility test}

When all six models (three with classic and three with aug) are run through the feasibility test, both Opt Ensemble and Opt GPR pass the test. In contrast, the classic NN failed the test, predicting a negative pressure from the augmented data. Despite the fact that five out of six models predicted only positive values.

\subsubsection{Testing the ground truth}
\begin{figure}[!htpb]
    \centering
    \subfloat[Opt Ensemble]{\includegraphics[width=7.7cm, height=6cm]{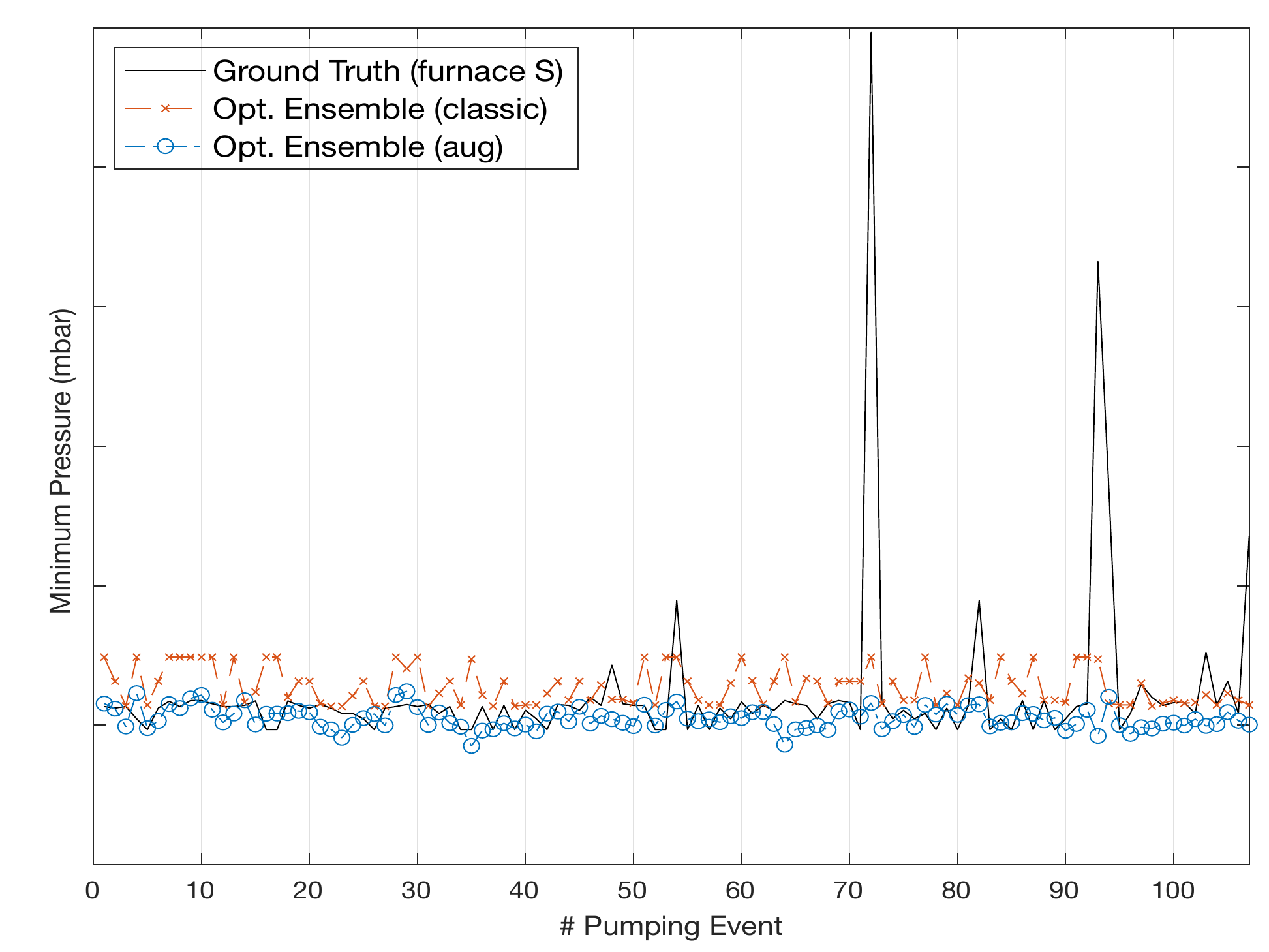}}\\
    \subfloat[NN]{\includegraphics[width=7.7cm, height=6cm]{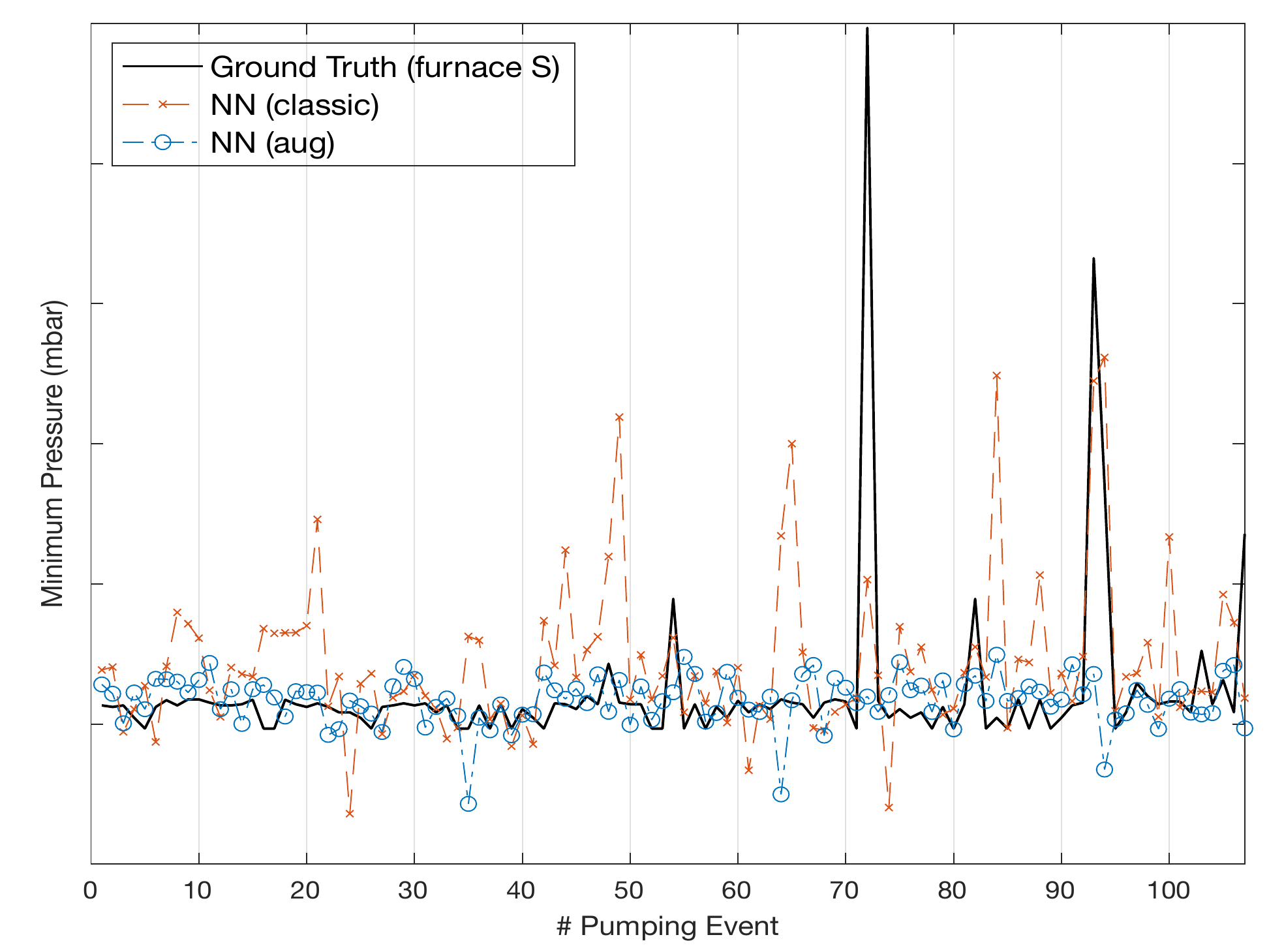}}
    \caption{A figure showing the prediction of Opt Ensemble and NN models with both classic and aug training.}
    \label{fig:resS}
\end{figure}

If a model passes the feasibility test, this scenario calculates the residual error with ground-truth data. With 60 seconds of input data, the residual error between the predicted minimum pressure and the actual minimum pressure is calculated. Table \ref{tab:test2} shows mixed MAE outcomes between furnace M and S individually between classic and aug training. However, there is an increase in the $R^2$ score and consistency in the results between bias and variance when trained with augmented samples. Figure \ref{fig:resS} illustrates the improvements, showing that training with augmented data surpasses the classical technique, in addition to spikes and deviations.

\subsubsection{Volume enclosed under a given residual error}
The last test estimates the volume of the trained model in temporal space whose residual MAE error is below the acceptable error threshold. For this experiment, a threshold of 1 mbar determined by Uddeholm field experts is selected. When the data is collected once per minute, we have 60 temporal dimensions with the first minute of pumping data as input.

Table \ref{tab:test3} shows that, compared to classic models, the enclosed capacity for aug models has increased dramatically. However, all six models cover less than 0.11\% of the whole augmented sample space, necessitating further research into a deep learning model for vacuum pumping predictions.

\begin{table}[!htpb]
\centering
\caption{Volume enclosed by augmented samples for 60 seconds of pumping data whose prediction residual error is below the threshold of 1 mbar. The volume of augmented data in this space is 1.1331e-17.}
\begin{tabular}{|c|c|c|} \hline
\textbf{Model} & \textbf{Volume} & \textbf{Increase \%}  \\ \hline
Opt Ensemble (classic) & 3.46e-45 & \\
Opt Ensemble (aug) & 6.39e-31 & 1.85e+16 \\ \hline
NN (classic) & 9.86e-52 & \\
NN (aug) & 1.93e-33 & 1.96e+20 \\ \hline
Opt GPR (classic) & 1.25e-35 & \\
Opt GPR (aug) & 7.48e-21 & 5.98e+16 \\ \hline
\end{tabular}

\label{tab:test3}
\end{table}

\subsubsection{The Test Oracle}

Finally, the output from the feasibility test and the findings from Tables \ref{tab:test2} and \ref{tab:test3} are input into the test oracle to measure the robustness of the six ML models under test. The thresholds for the pass-fail criteria for each sub-oracle will vary according to the application's sensitivity. For this experiment, we used the following thresholds determined by field experts: MAE of 1.5 mbar, $R^2$ threshold of 0.8, $\ell_\infty$ threshold of 25 mbar, and volume below the residual error threshold of 1.0e-35. Two of the six models pass the three sub-oracles and the main oracle, Opt Ensemble and NN, were both trained using augmented samples. And, based on the volume enclosed, Opt Ensemble (aug) covers $\approx$330x more volume than NN (aug), making Opt Ensemble trained with augmented samples the most robust of the six models.

\section{Conclusion}\label{conclusion}
In this paper, we show that the traditional 80-20 split between training and testing in software testing is inadequate for machine learning with small samples. Furthermore, this paper improves on the state-of-the-art software testing practices in robustness testing and presents a data augmentation technique that comprises newly developed test scenarios and oracles to determine a trained model's robustness in the absence of sufficient training and test data. The early experiments suggest that adding augmented data generated by the proposed approach to ML models' training improves the test data prediction accuracy and, more importantly, reduces the bias and variance difference between different vacuum pumping production data for ESR applications. Additionally, we identified that the six ML models under test do not capture events where the minimum pressure deviates substantially from the training distribution mean, which will be investigated in future work and require the development of a bespoke model for vacuum pumping application. The augmentation and testing approach is designed to aid future model development and how to design a robust testing methodology when working with small sample ML.

In the future, further model investigation will need to be carried out, including research into deep learning and the integration and use of several trained models simultaneously. This paper presents our best practice to date in addressing quality assurance challenges in an MLOps workflow. Although the paper focuses on vacuum pumping in the steel production industry, robustness testing methodology and general quality assurance workflow can be applied across multiple disciplines.

\begin{acks}
This work has been funded by the Knowledge Foundation of Sweden (KKS) through the Synergy Project AIDA - A Holistic AI-driven Networking and Processing Framework for Industrial IoT (Rek:20200067).
\end{acks}

\bibliographystyle{ACM-Reference-Format}
\bibliography{sample-base}

\end{document}